\documentclass[aps,prl,groupedaddress,twocolumn,showpacs,preprintnumbers,amsmath,amssymb,floatfix,superscriptaddress,bibnotes]{revtex4}

\pdfoutput=1
\usepackage{graphicx}
\usepackage{dcolumn}
\usepackage{bm}
\usepackage{color}
\usepackage{amsmath}
\usepackage{multirow}
\usepackage{natbib}

\newcommand{\Sn}{SnMo$_6$S$_8$}

\begin{document}

\title{\boldmath Real-space vortex glass imaging and the vortex phase diagram of {\Sn}}

\author{A.P. Petrovi\'c}
\affiliation{DPMC-MaNEP, Universit\'e de Gen\`eve, Quai Ernest-Ansermet 24, 1211 Gen\`eve 4, Switzerland.}
\author{Y. Fasano}
\altaffiliation[Present address:~]{Instituto Balseiro and Centro At\'omico Bariloche, Avenida Bustillo 9500, Bariloche, Argentina.}
\affiliation{DPMC-MaNEP, Universit\'e de Gen\`eve, Quai Ernest-Ansermet 24, 1211 Gen\`eve 4, Switzerland.}
\author{R. Lortz}
\affiliation{Department of Physics, The Hong Kong University of Science \& Technology, Clear Water Bay, Kowloon, Hong Kong.}
\author{C. Senatore}
\affiliation{DPMC-MaNEP, Universit\'e de Gen\`eve, Quai Ernest-Ansermet 24, 1211 Gen\`eve 4, Switzerland.}
\author{A. Demuer}
\author{A.B. Antunes}
\author{A. Par\'e}
\affiliation{Laboratoire des Champs Magn\'etiques Intenses CNRS, 25 rue des Martyrs, B.P. 166, 38042 Grenoble Cedex 9, France.}
\author{D. Salloum}
\author{P. Gougeon}
\author{M. Potel}
\affiliation{Sciences Chimiques, CSM UMR CNRS 6226, Universit\'e de Rennes 1, Avenue du G\'en\'eral Leclerc, 35042 Rennes Cedex, France.}
\author{\O. Fischer} 
\affiliation{DPMC-MaNEP, Universit\'e de Gen\`eve, Quai Ernest-Ansermet 24, 1211 Gen\`eve 4, Switzerland.}

\date{\today}

\begin{abstract}

Using scanning tunnelling microscopy at 400~mK, we have obtained maps of around 100 vortices in {\Sn} from 2~-~9\,T.  The orientational and positional disorder at 5 and 9\,T show that these are the first large-scale images of a vortex glass.  At higher temperature a magnetisation peak effect is observed, whose upper boundary coincides with a lambda anomaly in the specific heat.  Our data favour a kinetic glass description of the vortex melting transition, indicating that vortex topological disorder persists at fields and temperatures far below the peak effect in low-$T_c$ superconductors.  
  
\end{abstract}


\maketitle

Twenty years after the first attempt to produce a generalised vortex phase diagram for type-II superconductors, there is still no general consensus on this subject.  In particular, confusion remains over the nature and topology of the ``vortex glass'' phase~\cite{Fisher-1991,Giamarchi-1995,Mikitik-2001} and its relation to the peak effect observed in DC magnetisation and AC susceptibility of numerous type-II systems.  It has been claimed that the peak effect in low-$T_c$ materials is associated with the transition from a Bragg glass to a vortex glass~\cite{Banerjee-2001}.  Other peak effect interpretations include the elastic lattice softening model~\cite{Larkin-1979} and a multi-dynamic vortex liquid scenario~\cite{Higgins-1996}.  However, recent studies of Nb$_3$Sn do not give any indication of a phase transition from a Bragg glass to an intermediate disordered state within the peak regime: instead, the peak effect is interpreted as arising from the metastability of an underlying first-order vortex melting transition~\cite{Lortz-2006,Lortz-2007}.  It may therefore be considered as a zone dominated by strong thermal fluctuations and consequentially enhanced pinning.  

Extended real-space vortex imaging is the best method of clarifying the extent of disorder in the ($H$,$T$) phase diagram.  However, it remains a considerable challenge, with the inherent difficulties varying with the choice of superconductor.  In low-$T_c$ materials the disordered vortex phase typically spans a narrow window of phase space, thus limiting experimental accessibility~\cite{Lortz-2006}.  Images of the disordered phase have only been obtained in NbSe$_2$, where magnetic decoration reveals static disorder~\cite{Fasano-2002} and scanning tunnelling microscopy (STM) in the peak effect regime displays a crossover from collective vortex motion to positional fluctuations~\cite{Troyanovski-2002}.  In contrast, disorder occupies a far greater portion of phase space in high-$T_c$s due to their small coherence volumes and hence increased influence of thermal fluctuations.  Unfortunately high vortex mobility severely complicates the detection of any stable high-field vortex solid in these compounds~\cite{Fischer-2006}.  

The Chevrel phase {\Sn} is an attractive system to investigate since its extremely short coherence length $\xi \sim$~3~nm lies close to those of the high-$T_c$s, suggesting that disordered zones of its phase diagram may be more extensive and easily experimentally accessible.  An additional advantage is its quasi-3D crystal structure which should help to stabilise any disordered solid phase against melting at non-zero temperature~\cite{Fisher-1991}.  In this Letter, we present extended ($\sim$~100 vortices) STM images of a stable vortex glass at 400~mK in {\Sn}, far below the ($H$,$T$) range where the magnetisation peak effect is observed. The vortex glass has a high defect density, resulting in short-range positional and orientational order.  This contrasts with the defect-free quasi-long-range ordered low-field (2\,T) structure, suggestive of a Bragg glass.  Concerning the impact of the peak effect on the phase diagram, we observe a small lambda anomaly superimposed on the electronic specific heat jump at $T_{c2}(H)$.  This implies that the enhanced pinning within the peak effect region is due to the increasing influence of thermal fluctuations prior to a vortex melting transition~\cite{Lortz-2007}.  

\begin{figure}[bbb]
\centering
\includegraphics [width=\columnwidth] {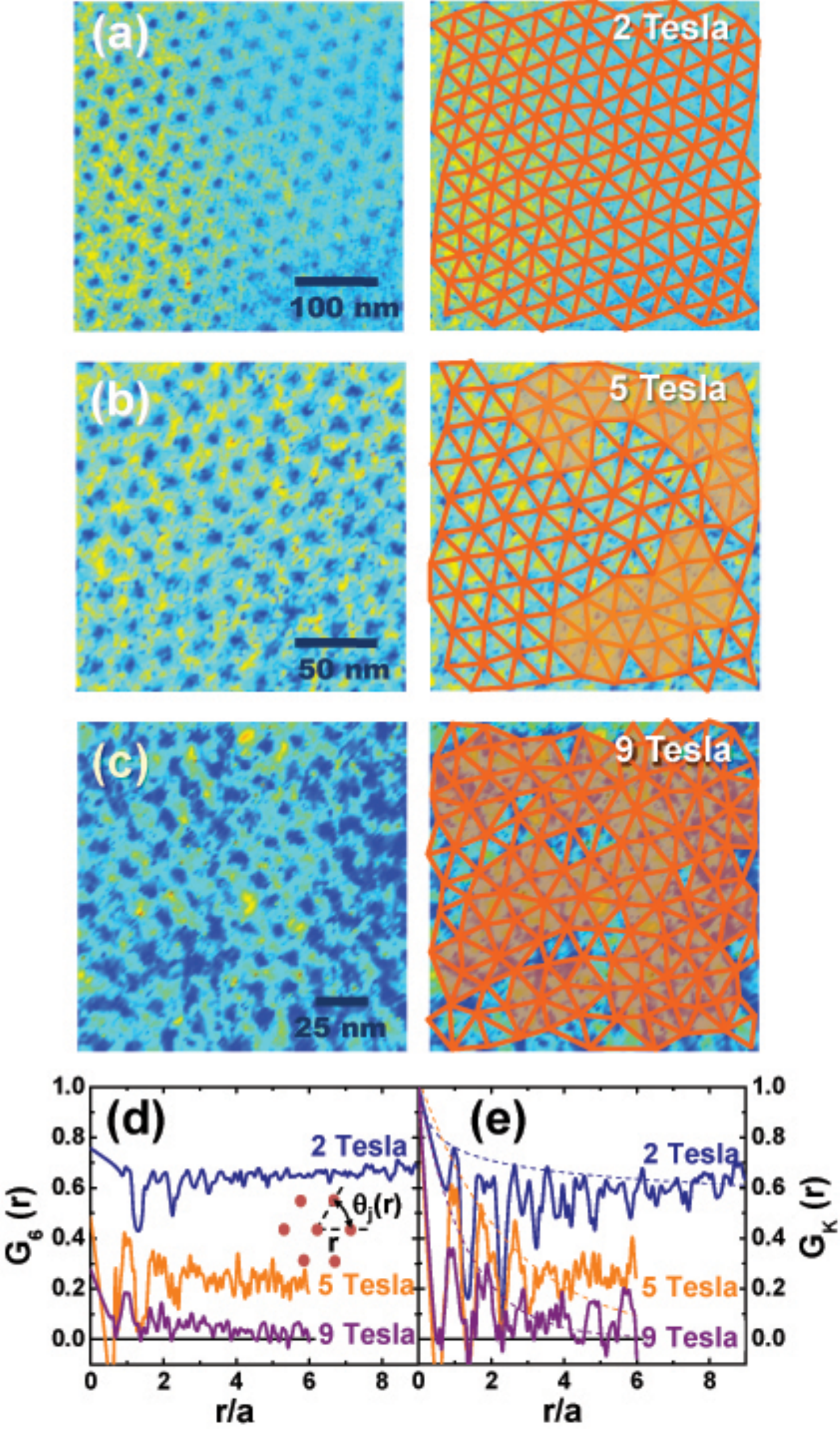}
\caption{\label{Fig_1} Vortex structure in {\Sn} at 400\,mK imaged by STM at (a) 2, (b) 5  and (c) 9\,T.  Right panels: corresponding Delaunay triangulations with nearest-neighbors linked by orange lines and topological defects highlighted with orange triangles. (d) Orientational correlation functions of the structures and a schematic of the angles considered for the calculation (see text). (e) Positional correlation functions and fits of the envelope using power-law (blue-dashed line, 2\,T) and exponential (orange and purple-dashed lines, 5 and 9\,T) decays.  The maps were acquired with a 60~M$\Omega$ junction using a lock-in technique~\cite{Fischer-2006}.}  
\end{figure}

Single crystals of {\Sn} (typical size 3.5~mm$^3$) were grown in sealed molybdenum crucibles at 1550~$^\circ$C.  Their high purity was verified by X-ray diffraction and AC susceptibility measurements, yielding $T_c$ = 14.2~K with an unprecedentedly low transition width of 0.1~K.  Magnetisation measurements were performed in a Quantum Design SQUID and a Lakeshore 7300 vibrating sample magnetometer (VSM), while heat capacity measurements in fields up to 28T were carried out in a high-resolution micro-calorimeter using the ``long-relaxation technique''~\cite{Lortz-2007-2}.  The low-temperature vortex structure (VS) was imaged using cleaved samples in a home-built high-vacuum ($\sim$10$^{-8}$~mbar) helium-3 STM operating in spectroscopic mode at 400~mK.  Samples were field-cooled from 18~K to 1.8~K at roughly 14~K/hour, kept at 1.8~K for 12 hours, then cooled to 400~mK and held for a further 36 hours prior to measurement.  This procedure allows the VS to relax towards its low-temperature equilibrium state.  Vortices were imaged as local minima in $\sigma_{pk}/\sigma_{ZBC}$ conductance-contrast maps, with $\sigma_{pk}$ the conductance at the Meissner state coherence peak energy (3~meV) and $\sigma_{ZBC}$ the zero-bias conductance~\cite{Fischer-2006}. Figure~\ref{Fig_1} shows the vortex positions (blue spots) for applied fields of 2, 5 and 9\,T on an atomically flat surface.  In contrast with previous reports on STM vortex imaging typically showing a few dozen vortices \cite{Fischer-2006,Sosolik-2003}, our images contain significantly more vortices ($\sim$~100).  This allows us to accurately study the field-evolution of the VS in {\Sn}.

The topological properties of the {\Sn} VS change remarkably with field: the 2\,T VS is almost perfectly hexagonal whereas the 5 and 9\,T VS are much more disordered.  This is evident from analysing the Delaunay triangulation \cite{Fasano-2008} of the structures, a construction allowing the identification of nearest-neighbors and hence topological defects (non-six-fold coordinated vortices). The 2\,T VS lacks defects whereas the high-field VS (5 and 9\,T) present numerous dislocations (pairs of five and seven-fold coordinated vortices highlighted with orange triangles), their density increasing with field.

A more quantitative description of the topology of the VS is provided by the orientational and positional correlation functions~\cite{Fasano-2008,Halperin-1978}. Figure~\ref{Fig_1}(d) shows the orientational correlation function $G_{6} (r)=\,<$$\Psi_{6}(0)\Psi_{6}^{\star}(r)$$>$ measuring the spatial evolution of the orientational order parameter $\Psi_{6}(r)= (1/n) \Sigma_{j=1}^{n} \exp{6i\theta_{j}(r)}$ \cite{Halperin-1978}, a quantification of the angular misalignment of vortices with respect to the principal directions of a perfect triangular lattice.  Within the field of view the 2\,T VS presents quasi-long-range orientational order, whereas in higher fields the orientational order is short-ranged: for large $r/a$, $G_{6} (r)$ at 5\,T is roughly 5 times smaller than that at 2\,T and $G_{6} (r)$ at 9\,T lies close to zero (for perfect orientational order $G_{6} (r) = 1$).  The lower degree of orientational order in the high-field VS is mainly associated with the proliferation of dislocations. Figure~\ref{Fig_1}(e) shows the spatial evolution of $G_{K}(r)$, the average of the positional correlation functions $G\mathbf{_{k_{i}}}(\mathbf{r})$ evaluated in the three principal directions of the vortex structure ($\mathbf{k_{i}}$ are obtained from the peak positions in the VS Fourier transform). Each $G\mathbf{_{k_{i}}}(\mathbf{r})=\,<$$\Psi\mathbf{_{k_{i}}}(0)\Psi\mathbf{_{k_{i}}}^{\star}(\mathbf{r})$$>$, where the positional order parameter $\Psi\mathbf{_{k_{i}}}(\mathbf{r})= \exp{i\mathbf{k_{i}}\cdot\mathbf{r}}$, measures the spatial evolution of the vortex displacements with reference to the sites of a perfect hexagonal lattice \cite{Fasano-2008,Halperin-1978}. In the 2\,T VS the envelope of $G_{K}(r)$ exhibits a power-law decay, a dependence consistent with the quasi-long-range positional order characteristic of Bragg glasses \cite{Nattermann-1990,Giamarchi-1994}.  However, for the high-field VS the envelope of $G_{K}(r)$ is better described by an exponential decay, indicative of short-range positional order. The positional order deteriorates with increasing field: for the 9\,T VS $G_{K}(r)$ decays faster and tends to a value close to zero at large $r/a$.  The considerable number of dislocations present in the high-field VS at temperatures as low as 400\,mK contrasts strongly with the defect-free 2\,T VS, indicating that the 5 and 9\,T VS depict a vortex glass~\cite{Fisher-1991} whilst the 2\,T image most probably represents a Bragg glass~\cite{Nattermann-1990,Giamarchi-1994}.  In order to confirm the latter, shaking experiments should be performed: this is not possible with our current experimental setup.  Since the vortex interaction energy grows in increasing magnetic field, the observation of a less ordered 5\,T VS suggests that a disorder-induced topological transition takes place in {\Sn} between 2 and 5\,T.  

\begin{figure}[bbb]
\centering
\includegraphics [width=8.5cm,clip] {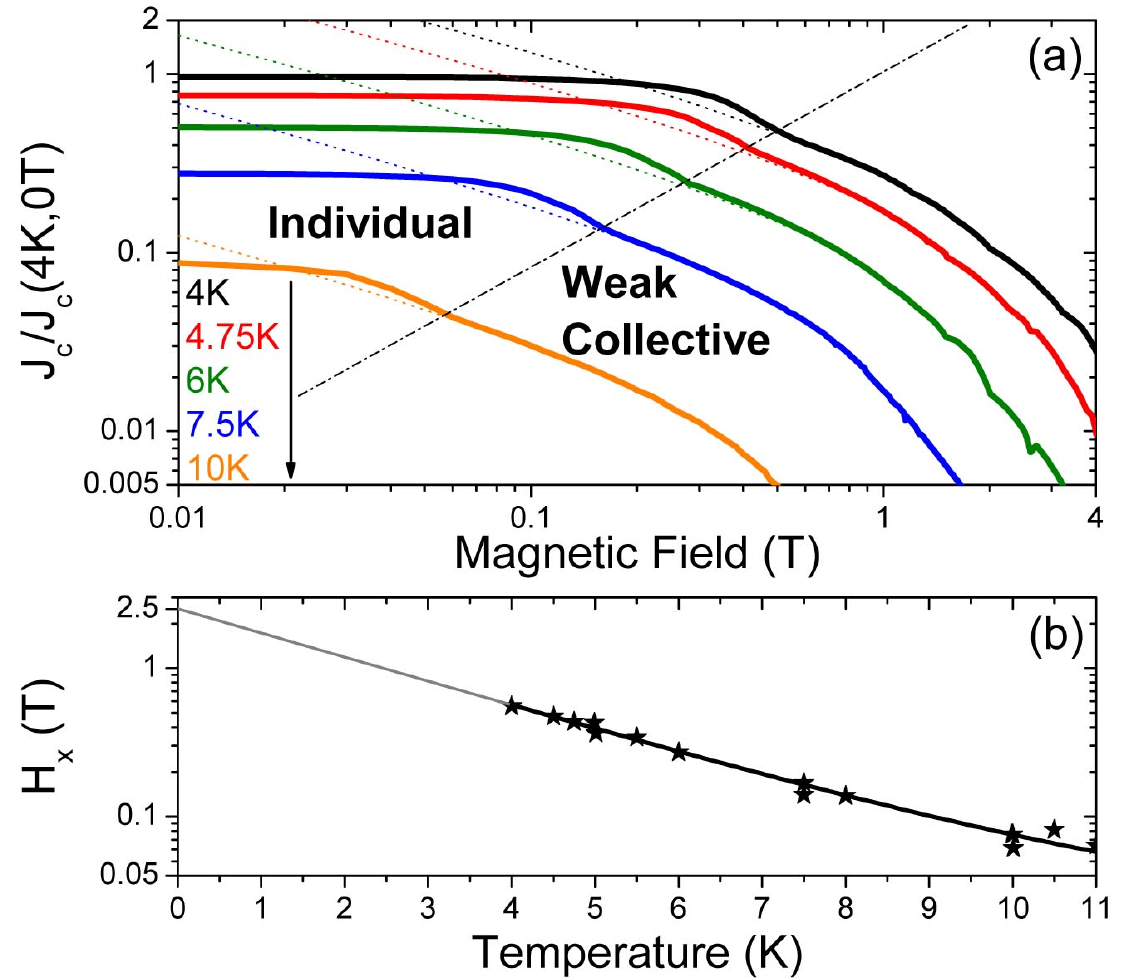}
\caption{\label{Fig_2} (a) Critical current versus field in {\Sn} (solid lines) obtained from magnetisation loops at different temperatures.  The crossover from individual to collective vortex pinning is identified as a kink in the curves.  Dashed lines: visual guides extrapolating the high-field $J_{c}(H)$.  (b) Pinning crossover field $H_{x}(T)$.  An exponential decay $ae^{-bT} + c$ extrapolates the fit to $T = 0$.}  
\end{figure}

\begin{figure}[bbb]
\centering
\includegraphics [width=8.5cm,clip] {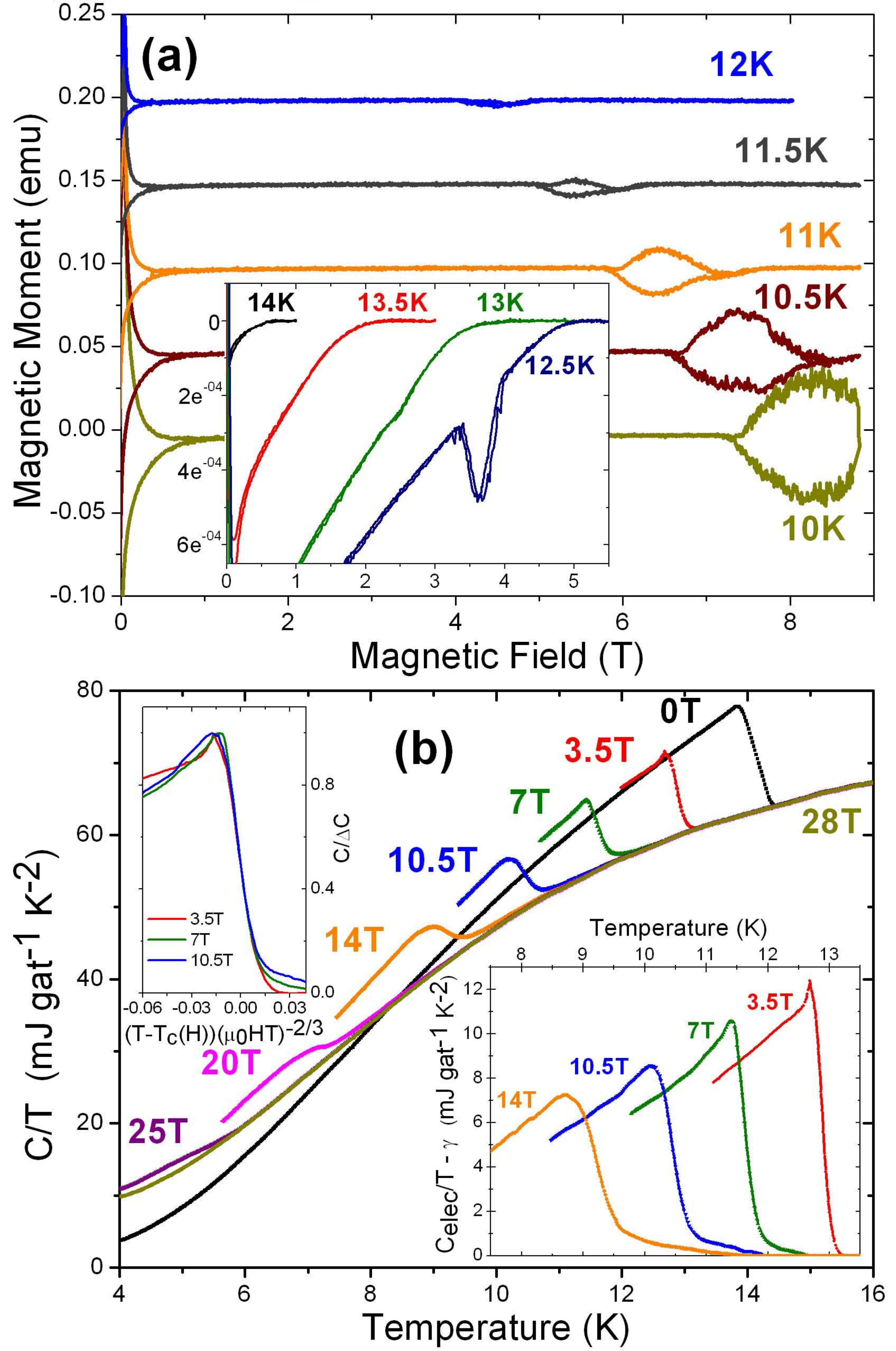}
\caption{\label{Fig_3} (a) Magnetisation hysteresis loops at $T$ =  10-12.5~K using a VSM.  Inset: Hysteresis loops from 12.5-14~K using the more sensitive SQUID.  (b) Total specific heat in {\Sn} from 0-28~T (the sample remains in the normal state at 28T above 4~K).  The superconducting transition temperature $T_{c2}(H)$ is defined at the midpoint of the heat capacity jump.  Right inset: Electronic specific heat $C_{elec}$ = $C$($H$)/$T$ - $C$(28T)/$T$ for 3.5-14~T.  Left inset: 3D-LLL scaling (see text) of $C_{elec}$ for 3.5-10.5~T.}
\end{figure}

It is instructive to examine the nature of vortex pinning within the phase space probed by STM, since in certain cuprates the low-temperature vortex order-disorder transition accompanies a crossover from individual to collective pinning~\cite{Piriou-2008}.  In Fig.~\ref{Fig_2}(a) we plot the evolution of the critical current density $J_{c}(H,T)$ (estimated from SQUID magnetisation loops).  Two zones are immediately visible: a low-field region where $J_c$ is roughly field-independent (characteristic of individual vortex pinning~\cite{Blatter-1994}) and a high-field region in which $J_c$ decreases with increasing field.  A distinct kink separates the two regions and we identify this as the crossover field $H_x$ between individual and weak collective pinning.  Although we are not able to measure our sample magnetisation at 400~mK, $H_{x}(T)$ is well-fitted by a simple exponential decay (Fig.~\ref{Fig_2}(b)).  Extrapolating this fit, we estimate $H_x$~$\sim$~2.2~$\pm$~0.30~T at 400~mK.  One possible explanation for our STM-imaged order-disorder transition between 2 and 5~T could therefore be the crossover from individual to weak collective pinning.  However, we stress that there is no simple relationship between pinning strength and topology due to the different lengthscales governing each property: the pinning potential ranges over $\sim~\xi$, whereas topological disorder is exhibited over lengthscales of the order of the vortex lattice constant $a \gg \xi$.

Order-disorder transitions in solid vortex matter are typically indicated by a peak effect, i.e. a jump in $J_c$ due to enhanced pinning, accompanied by a strongly hysteretic zone in the sample magnetisation~\cite{Paltiel-2000}.  Compared with the cuprates, the peak effect in low-$T_c$ superconductors is generally located at much higher fields and temperatures just below $T_{c2}(H)$.  Since $\xi$ in {\Sn} lies between those found in low-$T_c$s and cuprates it is important to verify the existence and location of any peak effect.  Figure~\ref{Fig_3}(a) displays magnetisation loops from 10~-~14~K: a clear peak effect may be seen below 14~K, broadening and moving linearly to higher fields as the temperature is reduced.  $H_{x}(T)$ is located at far lower fields than this local hysteresis, i.e. the peak effect and the order-disorder transition imaged by STM are two separate phenomena.  

In common with most experimental probes, magnetisation data are sensitive to irreversible contributions from flux pinning.  Conclusive information on the nature of any phase transition(s) underlying the peak effect may therefore only be obtained from a purely thermodynamic quantity, such as the specific heat.  In Fig.~\ref{Fig_3}(b) we show both the total ($C$) and electronic ($C_{elec}$) heat capacities of {\Sn} close to $T_{c2}(H)$.  It should be noted that there is no evidence for any phase transition in the field or temperature ranges compatible with the onset of the magnetisation peak effect.  However, at low fields a small lambda anomaly superimposed on the jump at $T_{c2}(H)$ is clearly visible, which is broadened and smeared out at higher fields.  The transition and anomaly exhibit good 3D lowest-Landau-level (3D-LLL) scaling~\cite{Tesanovic-1999}, as expected for the field range measured, thus confirming the fluctuation origin of the anomaly.  The situation in {\Sn} appears identical to that in Nb$_3$Sn~\cite{Lortz-2006,Lortz-2007}, where a similar lambda anomaly has been shown to be representative of a metastable first order vortex melting transition.  We therefore identify the peaks in $C/T$ as approximate vortex lattice melting temperatures $T_{VM}$.  

\begin{figure}[bbb]
\centering
\includegraphics [width=8.5cm,clip] {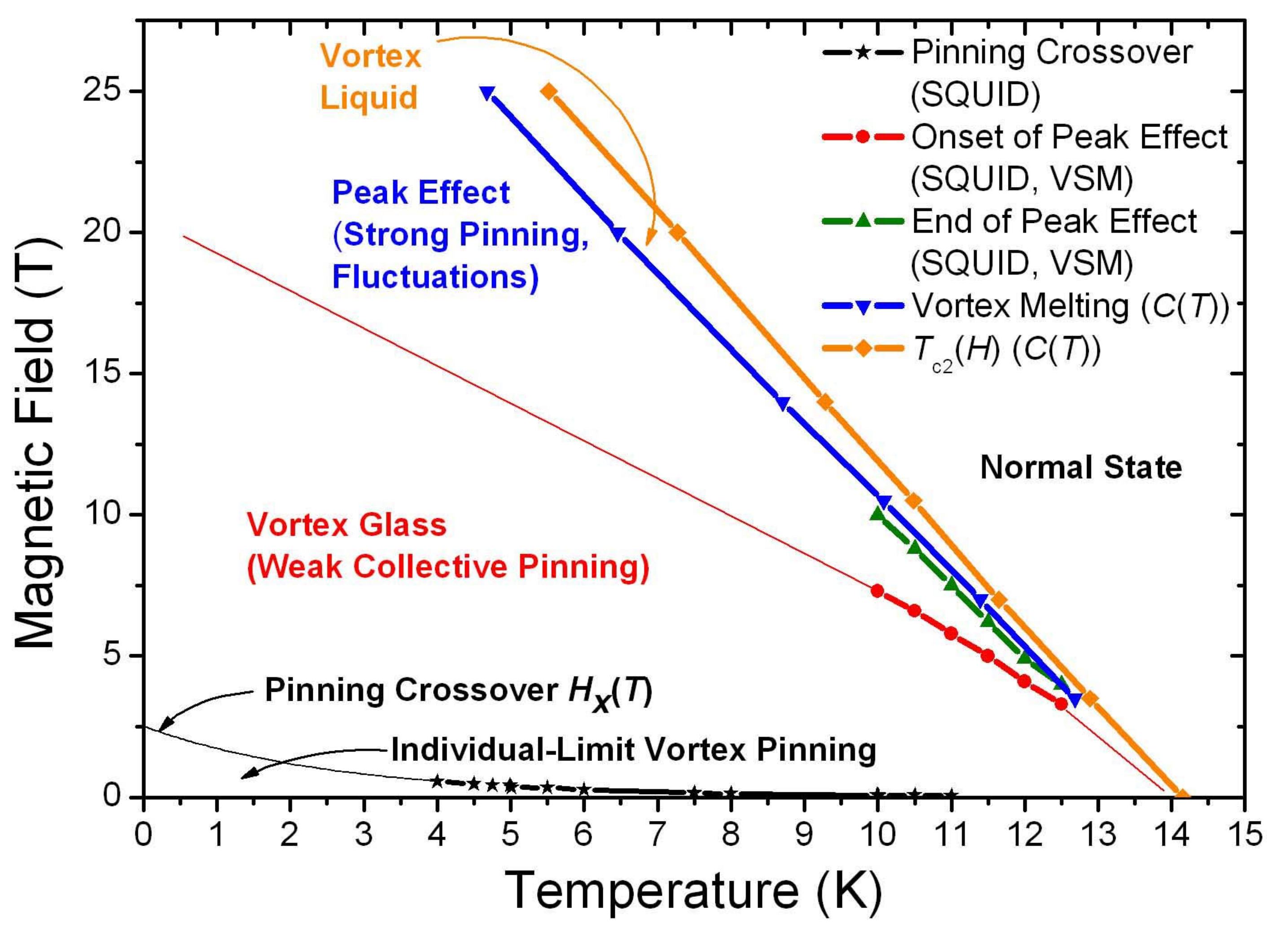}
\caption{\label{Fig_4} Vortex phase diagram of {\Sn}.  Measured data are displayed by points, extrapolated zones with thin lines.}  
\end{figure}

Combining data from bulk and local probes, we summarise the {\Sn} vortex phase diagram in Fig.~\ref{Fig_4}.  The majority of phase space is occupied by a vortex glass, a remarkable result which at first glance is unexpected for a low-$T_c$ material.  However, our specific heat data provides an explanation for this prevalence of disorder.  The absence of any latent heat (which would manifest itself as a spike in $C/T$ at $T_{VM}$) implies that a kinetic glass transition takes place: the vortex liquid has been undercooled and frozen into a disordered solid (the vortex glass).  This disorder persists beyond the peak effect region (which is merely the zone in which fluctuations from the melting transition enhance the pinning strength) down to low temperatures, except in the low-field limit where a quasi-ordered lattice (the Bragg glass) is stable.  The defect-free structure observed at 2\,T and 400~mK presents a positional order consistent with a Bragg glass, whilst the positionally and orientationally disordered high-field VS clearly indicate a vortex glass.  The possibility that the order-disorder transition between low and high-field phases coincides with the crossover from individual to collective pinning deserves further investigation.  In the event that these two phenomena are linked, both the topology of the vortex glass and $H_{x}(T)$ are expected to vary with the speed of undercooling through $T_{VM}$.  Ideally, future experiments should compare samples with different thermal histories to accurately probe the limits of disorder in the {\Sn} phase diagram.

\end{document}